\documentstyle[12pt]{article}
\begin{document}
\renewcommand{\thefootnote}{\fnsymbol{footnote}}
%                                       \begin{titlepage}
\begin{flushright}
MPI-Ph/93-78
\end{flushright}
%\vskip0.8cm
\begin{center}
{\LARGE
Deep Inelastic Scattering Beyond Perturbation Theory
            \\ }
\vskip1cm
 {\Large V.M.~Braun} $\footnote { On leave of absence from
St.Petersburg Nuclear
Physics Institute, 188350 Gatchina, Russia}$ \\
\vskip0.2cm
       Max-Planck-Institut f\"ur Physik   \\
       -- Werner-Heisenberg-Institut -- \\
        P.O.Box 40 12 12, Munich (Fed. Rep. Germany)\\
\vskip1cm
{\Large Abstract:\\}
\parbox[t]{\textwidth}{
I discuss possibilities to observe
the instanton-induced contributions
to deep inelastic scattering which correspond to nonperturbative
exponential
corrections to the coefficient functions in front of parton
distributions of the leading twist.
}
\\ \vspace{1.5cm}
{\em Talk given at the Leipzig workshop "Quantum Field
Theoretical Aspects of High Energy Physics",
Kyffh\"auser, September 20--24, 1993}
\end{center}
%                                               \end{titlepage}
\newpage
\vskip0.5cm
{\bf\large 1.}\hspace{0.5cm}
The deep inelastic lepton-hadron scattering at large momentum
transfers $Q^2$ and not too small values of the Bjorken
scaling variable  $x = Q^2/2pq$ is studied in much detail and
presents a classical example for the application of perturbative
QCD. The celebrated factorization theorems allow one to separate
the $Q^2$ dependence of the structure functions in coefficient
functions $C_i (x,Q^2/\mu^2 ,\alpha_s(\mu^2))$ in front of parton
(quark and gluon) distributions of leading twist
$P_i (x,\mu^2, \alpha_s (\mu^2))$
\begin{equation}
   F_2(x,Q^2) = \Sigma_i C_i (x,Q^2/\mu^2 ,\alpha_s(\mu^2))
              \otimes
               P_i (x,\mu^2, \alpha_s (\mu^2)),
\end{equation}
where
 \begin{equation}
      C(x) \otimes P(x) = \int_x^1 \frac{dy}{y} C(x/y) P(y),
 \end{equation}
the summation goes over all species of partons, and $\mu$ is the
scale, separating "hard" and "soft" contributions to the cross
section. At $\mu^2 = Q^2$ the coefficient functions can be
calculated perturbatively and are expanded in power series in
the strong coupling
 \begin{equation}
 C(x,1,\alpha_s(Q^2)) = C_0(x)+\frac{\alpha_s(Q^2)}{\pi} C_1(x)
     +\left(\frac{\alpha_s(Q^2)}{\pi}\right)^2 C_2(x)+\ldots
 \label{cpert}
 \end{equation}
 whereas their evolution with $\mu^2$ is given by famous
 Dokshitzer-Gribov-Lipatov-Altarelli-Parisi equations.
Going over to a low normalization point $\mu^2 \sim 1 GeV$,
one obtains the structure functions expressed in terms of the
parton distributions in the nucleon at this reference scale.
The parton distributions
absorb all the information about the dynamics of large distances and
are fundamental quantities extracted from the experiment.
Provided the parton distributions are known, all the dependence
of the structure functions on the momentum transfer is calculable
and is contained in the coefficient functions $C_i$.
Corrections to this simple picture come within perturbation
theory from the parton distributions of higher twists and are
suppressed by powers of the large momentum $Q^2$.
These higher-twist
contributions are relatively well understood theoretically,
and, unfortunately, very poorly experimentally. I have nothing to say
about them in this talk.

The picture described above presents a part of the common wisdom
about hard processes in the Quantum Chromodynamics, and in a more
or less detailed presentation can be found in any textbook.
Less widely known, is the fact that from the theoretical point
of view this picture is not complete. An indication that some
contributions may be missing, comes
 from the asymptotic nature of the perturbative series
in (\ref{cpert}). This series is non-Borel-summable, which means
that any attempt to attribute  a quantitative meaning to the
{\em sum} of the series in (\ref{cpert}) would produce an
exponentially small imaginary part
 $\sim i\exp\{-\mbox{\rm const}\cdot \pi/\alpha_s(Q^2)\}$,
 which is to be cancelled by the imaginary part coming from
nonperturbative contributions. Thus, separation between perturbative
and nonperturbative pieces in the cross section as the ones
which contribute to the coefficient function and the parton
distribution, respectively, cannot be rigorous.
A modern discussion of the asymptotical properties of the
perturbation series in QCD can be found in \cite{MUaachen,zakh92}.

In addition to {\em imaginary} exponential corrections which must
cancel identically against the corresponding nonperturbative
contributions, the coefficient functions may acquire also {\em real}
exponential corrections, which potentially produce observable
effects. In this talk I shall report on recent results by
I. Balitsky and myself \cite{bal93a,bal93b}, indicating that these
corrections are indeed present.
 We have found that
the deep inelastic cross section indeed possesses
exponential contributions of the form
$F(x)\exp[-4\pi S(x)/\alpha_s(Q^2)]$, where $S(x)$, $F(x)$
are  certain
functions of Bjorken $x$, which we are able to calculate in a
certain kinematical domain.
 Since the experimental data
are becoming more and more precise, it is of acute interest to find a
boundary for a possible accuracy of the perturbative approach, which
is set
by nonperturbative effects. Our study has been fuelled by
recent findings of an enhancement of instanton-induced effects
at high energies
in a related problem of the violation of baryon number
in the electroweak theory \cite{ring90,matt92}.
In the case of QCD the instanton-induced effects could
turn out to be numerically
large at high energies,
despite the fact that they correspond formally to  contributions
of a very high fractional twist
$\exp(-4\pi S(x)/\alpha_s(Q^2)) \sim (\Lambda_{QCD}^2/Q^2)^{bS(x)}$.

To be precise, in this talk I
consider the contribution to the structure
functions coming from the instanton-antiinstanton pair.
Contributions of single instantons are only present
for higher-twist
terms in the light-cone expansion and are less interesting.
A Ringwald-type enhancement
\cite{ring90} of the instanton-induced cross sections at
high energy can compensate the extra semiclassical
suppression factor $\exp(-2\pi /\alpha)$ accompanying
 instanton-antiinstanton contributions
compared to single-instanton
ones.  In such case the $\bar I I$ terms become
the leading ones owing to a bigger
power of the coupling in the preexponent.

As it is well known, the instanton contributions in QCD are
in general infrared-unstable.  In a typical situation
integrations over the instanton
size are strongly IR-divergent. Our starting point is
the observation that
this problem does not affect calculation of instanton contributions
to the coefficient functions. Let us introduce for a moment an
explicit IR cutoff $\Lambda_{IR}$ to regularize the integrals over
the instanton size. Then the contribution of
the instanton-antiinstanton pair to the cross section can be
written schematically as
\begin{equation}
   Q^2\sigma(Q^2) \sim (\Lambda_{IR}/\Lambda_{QCD})^{2b}
           + (\Lambda_{QCD}/Q)^{2bS(x)}.
\label{IR}
\end{equation}
 The second term in (\ref{IR}) gives an IR-protected contribution.
It depends in a nontrivial way on the external large momentum
and is identified  unambiguously with a contribution to
the coefficient function. The first term contributes to the
parton distribution. To be precise, one should separate in the
first term the contributions coming from instanton sizes above
and below the reference scale $\mu$, and to add
the contribution of
small-size instantons to the coefficient function.
Schematically, one has in this way
\begin{equation}
(\Lambda_{IR}/\Lambda_{QCD})^{2b} =
(\mu/\Lambda_{QCD})^{2b}  +
\left[(\Lambda_{IR}/\Lambda_{QCD})^{2b}-(\mu/\Lambda_{QCD})^{2b}
\right].
\end{equation}
However, this reshuffling of the $Q^2$-independent contribution
between the coefficient and the parton distribution does not
affect the observable cross section. It is analogous to an
ambiguity in the separation between contributions to
the coefficient function
and to the parton distribution in perturbation theory, induced by
possibility to use
different regularization schemes (e.g.
$\overline{MS}$ instead of $MS$, etc.). Hence, we can concentrate on
contributions of the second type in (\ref{IR}), which
are IR-protected.

\bigskip
{\bf\large 2.}\hspace{0.5cm}
The distinction between the instanton-induced contributions
to the coefficient functions, which are given by convergent
integrals over the instanton size, and the contributions to
parton distributions, given in general by IR-divergent integrals,
becomes especially transparent for an example of the
cross section of hard gluon scattering from a real gluon, see Fig.1a,
 considered
in detail in \cite{bal93a}. This process is not directly
physically relevant (e.g., the cross section is not gauge
invariant), but it serves as a good toy model.
Following Zakharov \cite{zakh90}, we calculate the cross section
by means of the optical theorem.
 The trick is to evaluate the
contribution to the functional integral
coming from the vicinity of the instanton-antiinstanton
configuration in Euclidian space, and calculate the
cross section by the analytical continuation to Minkowski
space and by taking the imaginary part.
Each hard gluon is substituted by the Fourier
transform of the instanton field in the singular gauge
at large momentum, and brings in
the factor
\cite{andrei}
\begin{equation}
   A_{\nu}^I(q)  \simeq
    \frac{i}{\rm g} (\sigma_{\nu}\bar{q} - q_{\nu})
    \left\{\frac{8\pi^2}{Q^4}-(2\pi)^{5/2}
\frac{\rho^2}{2Q^2}(\rho  Q)^{-1/2} e^{-\rho Q} .
\right\}
\label{1i}
\end{equation}
The first term in (\ref{1i}) produces a power-like divergent
integral over the instanton size $\rho$.
\begin{equation}
   \sigma \sim \frac{1}{Q^2}\int d \rho \sigma(\rho),\hspace{1cm}
    \sigma(\rho) \sim \rho^b
\label{1aaa}
\end{equation}
 All dependence on the
hard scale comes in this case through the explicit power
of $Q^2$ in front of the divergent integral. This is  a typical
contribution to the parton distribution --- in the present case
to the probability to find a hard gluon within a soft gluon.
%*************************** Figure 1  ****************************
\begin{figure}
    \begin{center}
        \begin{picture}(70,70)
        \end{picture}
    \end{center}
    \caption[xxx]{ \small
                  The contribution of the instanton-antiinstanton pair
                   to the cross section
                   of hard gluon-gluon scattering (a),
                   structure function of a gluon (b) and of a quark
                   (c,d). Wavy lines are (nonperturbative) gluons.
                    Solid lines are
                   quark zero modes in the case that they are
                    ending at the instanton (antiinstanton),
                    and quark propagators at the $\bar I I$ background
                    otherwise.
                    }
   \label{pic.a}
\end{figure}
The second term gives rise to a completely different behavior.
The cross section is given in this case by the following
integral over the common scale of the instanton and antiinstanton
$\rho_I \sim \rho_{\bar I}$ and over their separation $R$ in the c.m.
frame \cite{khoze,bal93a}:
\begin{equation}
\int \rho\,d\rho\int dR_0\,\exp\left\{-2Q\rho+ER_0-
\frac{4\pi}{\alpha_s(\rho)}S(\xi)\right\}.
\label{1q}
\end{equation}
Three important ingredients in this expression are: the factor
$\exp(-2Q\rho)$, which comes from the two hard gluon fields,
the factor $\exp(ER_0)$, which is obtained from the standard
exponential factor $\exp[i(p+q)R], E^2 = (p+q)^2$ by the rotation
to Minkowski space, cf. \cite{zakh90}, and the action
 $S(\xi)$  evaluated on the instanton-antiinstanton
configuration. The normalization is such that $S(\xi)=1$ for
an infinitely separated instanton and antiinstanton, and $\xi$
is the conformal parameter \cite{yung88}
\begin{equation}
  \xi=\frac{R^2+\rho_1^2+\rho_2^2}{\rho_1\rho_2}.
\label{xi}
\end{equation}
Writing the action as a function of $\xi$ ensures that the
interaction  between instantons is small in two different limits:
for a widely separated $I \bar I $ pair, and for a small instanton
put inside a big (anti)instanton, which are related to each other
by the conformal transformation. In the limit of large $\xi$ the
expansion of $S(\xi)$ for the dominating
maximum attractive $I \bar I $ orientation reads \cite{yung88}
\begin{equation}
  S(\xi) =  \left(1-\frac{6}{\xi^2}+O(\ln(\xi)/\xi^4)\right)
\label{action}
\end{equation}
where the $1/\xi^2$ term corresponds to a slightly corrected
 dipole-dipole interaction.  Thus, the action  $S(\xi)$
decreases with the distance between instantons, so that
the instanton and the instanton effectively attract each
other. This attraction results in the exponential increase
of the cross section --- the effect found by Ringwald \cite{ring90}.
Further terms in the expansion of the action can be obtained by the
so-called valley method \cite{bal86}, and a typical solution
(conformal valley) \cite{yung88}
gives a monotonous function of the conformal parameter, which turns
to zero at $R\rightarrow 0$. In the traditional language, the
valley approach corresponds to the summation of all so called
"soft-soft" corrections arising from the particle interaction
in the final state. Main problem is in the evaluation of
 "hard-hard" corrections \cite{MU91}, which
come from particle interaction in the initial state.
These corrections are likely to decrease the cross section,
and in physical terms
must take into account an (exponentially small) overlap
between the initial state, which involves a few hard quanta, with the
semiclassical final state \cite{banks}.
Thus, the instanton-antiinstanton action is
substituted by an effective "holy grail" function, which
determines the leading exponential factor for the
semiclassical production at high energies, and
which received a lot of attention in recent years.
Unitarity arguments \cite{zakh91,magg91} suggest that
the decrease of the action will stop at values of order
$S(\xi)\simeq 0.5$. In a recent preprint \cite{DP93}
Diakonov and Petrov argue that $S(\xi)$ indeed decreases up
to the value
$1/2$ at a certain energy of order of the sphaleron mass, and
then starts to increase, so that the semiclassical production
cross section is resonance-like. The question seems to us
to be not settled finally. In this study, we have taken the
value $S=1/2$ as a reasonable guess for the residual suppression,
and assumed that the behavior of the "true" function $S(\xi)$
for $S(\xi)>1/2$ is close to that given by the conformal valley
\cite{yung88}. The latter assumption is supported by numerical
studies, e.g. in \cite{DP93}.

 To the semiclassical accuracy the integral in (\ref{1q})
is evaluated by a saddle-point method.
 The saddle-point equations take the form \cite{bal93a}
\begin{eqnarray}
 Q\rho_\ast &=&
\frac{4\pi}{\alpha_s(\rho_\ast)}(\xi_\ast-2)S'(\xi_\ast)
+bS(\xi_\ast)\,,
        \nonumber\\
E \rho_\ast=  &=&\frac{8\pi}{\alpha_s( \rho_\ast)}
\sqrt{\xi_\ast-2}\,S'(\xi_\ast)\,,
\label{saddleeq}
\end{eqnarray}
where $S'(\xi)$ is the derivative of $S(\xi)$ over $\xi$, and
$\rho_\ast, \xi_\ast$ are the saddle-point values for the
instanton size and the conformal parameter, respectively.

Neglecting in (\ref{saddleeq}) the terms proportional to
$b=(11/3) N_c - (2/3)n_f$, which come from the differentiation
of the running coupling (and produce a small correction),
one finds
\begin{eqnarray}
      \xi_\ast & = & 2 + \frac{R_\ast^2}{\rho^2_\ast}
        = 2 \frac {1+x}{1-x} ,
\nonumber\\
     Q\rho_\ast &=& \frac{4\pi}{\alpha_s(\rho_\ast)}
     \frac{12}{\xi_\ast^2}
\label{saddle}
\end{eqnarray}
A numerical solution of the saddle-point equations in (\ref{saddleeq})
for the particular expression of the action $S(\xi)$ corresponding
to the conformal instanton-antiinstanton
valley is shown in Fig.2.
%*************************** Figure 2  ****************************
\begin{figure}
    \begin{center}
          \begin{picture}(70,70)
          \end{picture}
    \end{center}
    \caption[xxx]{ \small
                  The non-perturbative scale in deep inelastic
                   scattering  (instanton size $\rho_\ast^{-1}$),
                  corresponding to the solution of saddle-point
                  equations in (\ref{saddleeq}) as a
                  function of $Q$ and for $S(\xi_\ast)\sim 0.5-0.6$
                  ($\xi_\ast\sim 3-4$).
 }
   \label{pic.b}
\end{figure}
Note that the difference between the hard scale $Q^2$ and the
effective scale for nonperturbative effects $\rho^{-2}_\ast$ is
numerically very large.  This is a new situation compared to
 calculations of instanton-induced
 contributions to two-point correlation functions, see
e.g. \cite{andrei,NSVZ80,DS80}, where the
size of the instanton is of order of the large virtuality.
The effect is
 that the instanton-induced contributions to deep
inelastic scattering may turn out to be non-negligible at the
values
$Q^2 \sim 1000 GeV^2$, which are conventionally
considered as a safe domain for perturbative QCD.

In the case of hard gluon-gluon scattering it is easy to collect
all the preexponential factors (to the semiclassical accuracy).
The result for the scattering of a transversely polarized hard
gluon from a soft gluon reads \cite{bal93a}
\begin{eqnarray}
2E^2\sigma_{\perp} & =&
\frac{4}{9} d^2 \frac{(1-x)^2+x^2}{x^2 (1-x)^2}
\pi^{13/2} \left(\frac{2\pi}{\alpha(\rho_\ast)}\right)^{21/2}
\exp \left[ -
\left(\frac{4\pi}{\alpha_s(\rho_\ast)}+2b\right)
S(\xi_\ast )\right]\,.
\label{gluon}
\end{eqnarray}
 It is expressed in terms of the saddle-point values of
  $\rho$ and $\xi$.
Here $d\simeq 0.00363$ (for $n_f=3$) is a constant which enters
the expression for the instanton density
\begin{equation}
 d  =  \frac{1}{2}C_1 \exp[n_f C_3-N_c C_2)],
\label{d}
\end{equation}
$C_1 = 0.466, C_2 = 1.54, C_3 = 0.153$ in the $\overline{MS}$
scheme.

Note that the preexponential factor in (\ref{gluon}) is
calculated to the leading semiclassical accuracy, i.e. in the
limit $ x\rightarrow 1$, while in the exponential factor
$\exp\{-(4\pi/\alpha_s +2b)S(\xi_\ast)\}$ we used
the expression for $S(\xi)$ suggested by the valley method,
which is likely to be a good approximation provided $S(\xi)>1/2$,
alias $x>0.25-0.3$, cf. (\ref{saddle}).
 With this restriction,
and at $\alpha_s(\rho_\ast) \simeq 0.3-0.4$,
the expression on the r.h.s. of (\ref{gluon}) reaches values of order
$10^{-2}-10^{0}$,
which means that at $Q^2 \sim 100-1000 GeV^2$ and $x< 0.25-0.40$
the nonperturbative contribution appears to be significant.

\bigskip
{\bf\large 3.}\hspace{0.5cm}
Similar contributions are present  in the structure functions
of deep inelastic lepton-hadron scattering, but the calculation
turns out to be much more involved \cite{bal93b}.
 The situation proves to be  somewhat
simpler for the case of deep inelastic scattering from a real
gluon. To this purpose we need to evaluate
\begin{eqnarray}
T_{\mu\nu}& =& i\int d^x\,e^{iqx} \langle A^a(p),\lambda |
T\{ j_\mu(x) j_\nu (0)\}|A^a(p),\lambda\rangle
\nonumber\\
W_{\mu\nu}&=&\frac{1}{\pi} \mbox{Im}\, T_{\mu\nu}=
         \\
&&\mbox{} =
\left(-g_{\mu\nu}+\frac{q_\mu q_\nu}{q^2}\right)F_L(x,Q^2)
+\left(\frac{p_\mu p_\nu}{pq}-
\frac{p_\mu q_\nu + q_\mu p_\nu}{q^2} +g_{\mu\nu}
\frac{pq}{q^2}\right) F_2(x,Q^2)
\nonumber
\label{str}
\end{eqnarray}
There are two technical problems to be solved. First of all, the
separation of the finite contribution to the coefficient function
under the background of a divergent contribution to the
parton distribution is no longer given by a simple formula
in (\ref{1i}). Instead, we extract the contribution of interest
by making an analytic continuation of integrals over the instanton
size $\rho$ from {\em negative} values of $\rho^2$.
For a typical integral we write down, e.g.
\begin{eqnarray}
\int_0^\infty d\rho^2 \frac{(\rho^2)^{\mu+n-1} \Gamma(\lambda)}
{(T^2+\rho^2)^\lambda} & = &
\frac{\Gamma(\lambda)}{2i \sin [\pi (\lambda-\mu-n)]}
\int_{-\infty}^0\!d\rho^2\,(-\rho^2)^{\mu+n- \lambda -1}
\nonumber\\
&\times &
\left[\left(\frac{\rho^2+ i\epsilon}{T^2+\rho^2+ i\epsilon}
\right)^\lambda -\mbox{c.c.}\right]
 = \frac{\Gamma(\lambda-\mu-n)\Gamma(\mu+n)}{(T^2)^{\lambda-\mu-n}}
\label{rhoint}
\end{eqnarray}
 The second, and main complication, comes from the necessity to
consider the quark propagator in the $\bar I I$ background,
see the diagram in Fig.1b.  Neglecting
corrections of order $\rho^2/R^2$ in the preexponential
factor, we can make use of the cluster expansion \cite{andrei},
and keep the first nontrivial  term only:
 $$
  \langle x|\nabla_{I\bar I}^{-2}\bar \nabla_{I\bar I}|0\rangle
 =
 \int dz\, \langle x|\nabla_1^{-2} \bar\nabla_1|z\rangle
 \sigma_\xi \frac{\partial}{\partial z_\xi}\langle z|
 \bar\nabla_2 \nabla_2^{-2} |0\rangle .
 $$
Here and below the subscript '1' refers to the antiinstanton
with the size $\rho_1$ and the position of the center
 $x_{\bar I} = R+T$,
and the subscript '2'  refers to the instanton with the size
$\rho_2$ and the center at $x_I = T$.
 We use   conventional notations $\nabla = \nabla_\mu\sigma_\mu$ and
$\bar\nabla = \nabla_\mu\bar\sigma_\mu$, etc,  where
$\sigma_\mu^{\alpha\dot\alpha} = (-i\sigma, 1)$,
$\bar\sigma_{\mu\dot\alpha\alpha} = (+i\sigma, 1)$, and
$\sigma$ are the standard  Pauli matrices.
The expressions for quark propagators at the
one-instanton (antiinstanton)
background are given in  \cite{brown}.

We have found that to the leading accuracy in the strong coupling
the instanton contribution in Fig.1b comes from the following
integration regions in the coordinate space:
\begin{eqnarray}
      z^2 &\sim & (z-x)^2  \sim  x^2
\nonumber\\
     (x-R-T)^2+\rho_1^2&\sim & T^2+\rho_2^2  \sim  x^2
 \nonumber\\
     (z-R-T)^2+\rho_1^2 &\sim &  (z-T)^2 +\rho_2^2\sim x^2/\alpha_s
\nonumber\\
     T^2 &\sim & R^2 \sim \rho_1^2\sim \rho_1^2 \sim  x^2/\alpha_s
\end{eqnarray}
We remind
  that all the calculation is done in Euclidian space, and the
evaluation of integrals by means of the analytical continuation
effectively corresponds to the integration over
negative values of $\rho^2$, see (\ref{rhoint}).

Hence the integration over $z$ in the cluster expansion of the
quark propagator can
 be done in
the "light-cone" approximation:
\begin{equation}
\int dz\, \frac{F(z)}{(x-z)^4 z^4} =
\frac{\pi^2}{x^4} \int^1_0
d\gamma \frac{F(\gamma x)}{\gamma (1-\gamma)}
 +O(\sqrt{\alpha_s}),
\end{equation}
where $F(z)$ is an arbitrary function containing all other
possible denominators like $(z-R-T)^2+\rho_1^2$ etc.
This is a major simplification compared to the general case.

After a considerable algebra,
we obtain the
following answer for the $\bar I I$ contribution
to the structure function of a real gluon:
\begin{eqnarray}
F_1^{(G)}(x,Q^2) &=&\sum_q e^2_q
\frac{1}{9\bar x^2}
\frac{ d^2\pi^{9/2}}{bS(\xi_\ast)[bS(\xi_\ast)-1]}
\left(\frac{16}{\xi_\ast^3}\right)^{n_f-3}
\nonumber\\
&&\mbox{}\times
\left(\frac{2\pi}{\alpha_s(\rho_\ast^2)}\right)^{19/2}
\!\!\exp\left[ -
\left(
\frac{4\pi}{\alpha_s(\rho_\ast^2)} +2b\right)
S(\xi_\ast)\right]
\label{answer}
\end{eqnarray}
where the expressions for $\rho_\ast$ and $\xi_\ast$ coincide
to the ones given in (\ref{saddle}).
To our accuracy, we find that the instanton-
induced contributions obey the Callan-Gross relation
$F_2(x,Q^2)= 2x F_1(x,Q^2)$.

The expression in (\ref{answer}) presents our main result.
It gives  the exponential correction to the coefficient
function in front of the gluon distribution of the leading twist
in (\ref{cpert}).
The exponential factor is exact to the accuracy of (\ref{action}).
The preexponential factor
is calculated to leading accuracy
in the strong coupling and up to corrections of order $O(1-x)$.
 The corresponding
contribution to the structure function of the nucleon is obtained
in a usual way, making a convolution of (\ref{answer}) with a
distribution of gluons in the proton at the scale $\rho_\ast^2$.

In the language of the operator product expansion,
the answer in (\ref{answer}) can be rewritten as the exponential
nonperturbative contribution to the high $n\sim \pi/\alpha_s$
moment of the structure function
\begin{equation}
   M_1^{(G)}(n,Q^2) = \int_0^1 dx\, x^n F_1^{(G)}(x,Q^2)
\label{moments}
\end{equation}
Taking $n = \hat n \cdot 4\pi/\alpha_s(\rho_\ast)$ with $\hat n$ of
order unity, one can evaluate the integral in (\ref{moments}) with
$F_1^{(G)}(x,Q^2)$ given in (\ref{answer}) by the saddle point
method. The saddle-point equation reads, approximately
\begin{equation}
   \hat n = S'(\xi_\ast)  (\xi_\ast^2-4)/4
\end{equation}
where $\xi_\ast$ is related to the saddle-point value of $x$ by
(\ref{saddle}). The lowest possible value $\xi_\ast\simeq 3 $,
corresponding within the valley-method approximation to
$S(\xi_\ast)\simeq 1/2$, $S'(\xi_\ast)\simeq 0.27$, thus yields
$\hat n \simeq 0.3$.  Hence the expression in (\ref{answer})
is applicable to the evaluation of instanton-induced  corrections
to the coefficient functions in front of local operators starting
from $n\simeq 10-12$.

Leading contribution to (\ref{answer}) in the perturbation theory
is due to the mixing with the flavor-singlet quark distribution
and is given by the box graph:
\begin{eqnarray}
F_1^{(G)}(x,Q^2)_{pert} =  \sum_q e^2_q
[x^2+\bar x^2]\frac{\alpha_s(Q^2)}{2\pi}
\ln\left[\frac{Q^2\bar x}{\mu^2 x}\right] ,
\label{box}
\end{eqnarray}
where in order to compare to the instanton
contribution in (\ref{answer}) one should choose the scale $\mu$
to be of order $\rho_\ast$.

The instanton-antiinstanton contribution to the structure function
of a quark contains a similar contribution shown in Fig.1c.
The answer reads
\begin{eqnarray}
F_1^{(q)}(x,Q^2) &=&
\left[\sum_{q'\neq q} e^2_{q'} +\frac{1}{2}e^2_q\right]
 \frac{128}{81\bar x^3}
\frac{d^2\pi^{9/2} }{bS(\xi_\ast)[bS(\xi_\ast)-1]}
\left(\frac{16}{\xi_\ast^3}\right)^{n_f-3}
\nonumber\\
&& \mbox{}\times
\left(\frac{2\pi}{\alpha_s(\rho_\ast^2)}\right)^{15/2}
\exp\left[ -
\left(
\frac{4\pi}{\alpha_s(\rho_\ast^2)} +2b\right)
S(\xi_\ast)\right]
\label{qanswer}
\end{eqnarray}
However, in this case additional contributions exist of the
type shown in Fig.1d. They are finite (the integral over
instanton size is cut off at $\rho^2 \sim x^2/\alpha_s$),
 but the relevant instanton-antiinstanton separation
 $R$ is small, of order $\rho$.
This probably means that the structure of nonperturbative
contributions to quark distributions is more complicated.
This question is under study.
The answer given in (\ref{qanswer}) presents the
contribution of the particular saddle point in (\ref{saddle}).

\bigskip
{\bf\large 4.}\hspace{0.5cm}
The instanton-induced contribution to the structure function of a
gluon in (\ref{answer}) is shown as a function of Bjorken x for
different values of $Q\sim 10-100 GeV$ in Fig.3.
The contribution of
the box graph in (\ref{box}) is shown by dots for comparison.
The low boundary for possible values of $Q$ is determined by
the condition that the effective instanton size is not too
large. At $Q=10 \,GeV$ we find $\rho_\ast \simeq 1\, GeV^{-1}$, cf.
Fig.2. This value is sufficiently small, so that instantons
are not distorted too strongly by large-scale vacuum fluctuations.
Another limitation is that the valley approach to the
calculation of the "holy grail" function $S(\xi)$ is likely to be
justified at $S(\xi) \ge 1/2 $, which
translates to the condition that $x>0.3-0.35$.
%*************************** Figure 3  ****************************
\begin{figure}[t]
    \begin{center}
\begin{picture}(100,120)
\end{picture}
    \end{center}
\caption[xxx]{  \small
                  Nonperturbative contribution to the structure function
                  $F_1(x,Q^2)$ of a real gluon (\ref{answer})
                  as a function of $x$ for
                   different values of $Q$ (solid curves).
                   The leading perturbative contribution (\ref{box})
                   is shown for comparison by dots. The dashed curves
                   show lines with the constant effective value of
                   the action on the $\bar I I$ configuration.
 }
   \label{pic.c}
\end{figure} \nopagebreak
Numerical results are strongly sensitive to the particular value
of the QCD scale parameter.
 We use the two-loop expression for the coupling
with three active flavors,
and the value $\Lambda_{\overline MS}^{(3)} = 290 MeV$ which
corresponds to  $\Lambda_{\overline MS}^{(4)} = 240 MeV$ \cite{PDT}.
The corresponding value of the coupling at the scale of $\tau$-lepton
mass is
$\alpha_s (m_\tau)= 0.29$.
The new ALEPH data \cite{ALEPH} indicate a somewhat higher value
$\alpha_s (m_\tau)= 0.33\pm 0.05$.
Since the dependence on the coupling is exponential, the 20\% increase
of $\alpha_s(\rho_\ast)$ induces the increase of the cross section
by an order of magnitude! Together with uncertainties in the function
$S(\xi)$ and in the preexponential factor, this indicates that the
particular curves given in Fig.3 should not be taken too seriously,
and rather give a target for further theoretical (and experimental?)
studies to shoot at.

To summarize, we have found that
 instantons produce a well-defined and calculable
contribution to the cross section of deep inelastic scattering
 for sufficiently large values of x and large $Q^2\sim 100 - 1000
GeV^2$,
which turns out, however, to be rather small ---
of order $10^{-2}-10^{-5}$ compared to the
perturbative cross section.
This  means that the
accuracy of standard perturbative analysis is sufficiently
high, and that there is not much hope to observe
the instanton-induced contributions to the total deep inelastic
cross section experimentally.
However, instantons are likely to produce events with
a very specific structure of the final state, and such
peculiarities may be subject to experimental search.
The dominating Feynman diagrams in our calculation correspond to
 $2\pi/\alpha(\rho_\ast)\sim 15 $ gluons and $2n_f-1 =5$ quarks
in the final state, with the low virtuality of order $\rho_\ast^{-1}
\sim 1\,GeV$, and which are produced in the S-wave in the c.m.
frame of
the colliding partons.
It is not likely that these can be resolved as minijets, and
we rather expect a spherically symmetric production of final state
hadrons. The effect is likely to be resonance-like, that is
present in  a narrow interval of values of Bjorken x of order
0.25--0.35 (in the parton-parton collision).
In any case, finding of an instanton-induced particle production
at high energies is a challenging problem, and further theoretical
efforts are needed to put it as a practical proposal to
experimentalists.

\bigskip
{\bf\large 5.}\hspace{0.5cm}
I would like to thank Ian Balitsky for a very rewarding collaboration.
It is a pleasure to
acknowledge also useful discussions with  M. Beneke, M.A. Shifman
and V.I. Zakharov.

%\clearpage

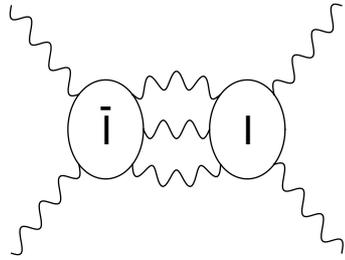
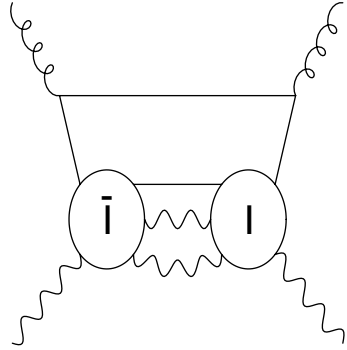
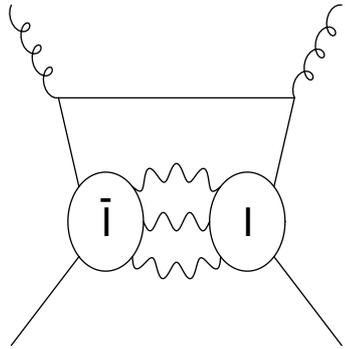
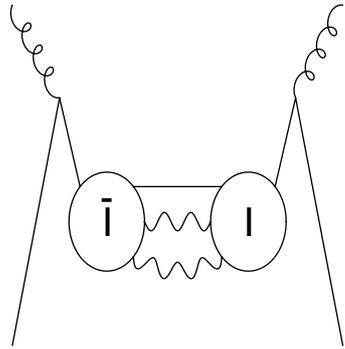
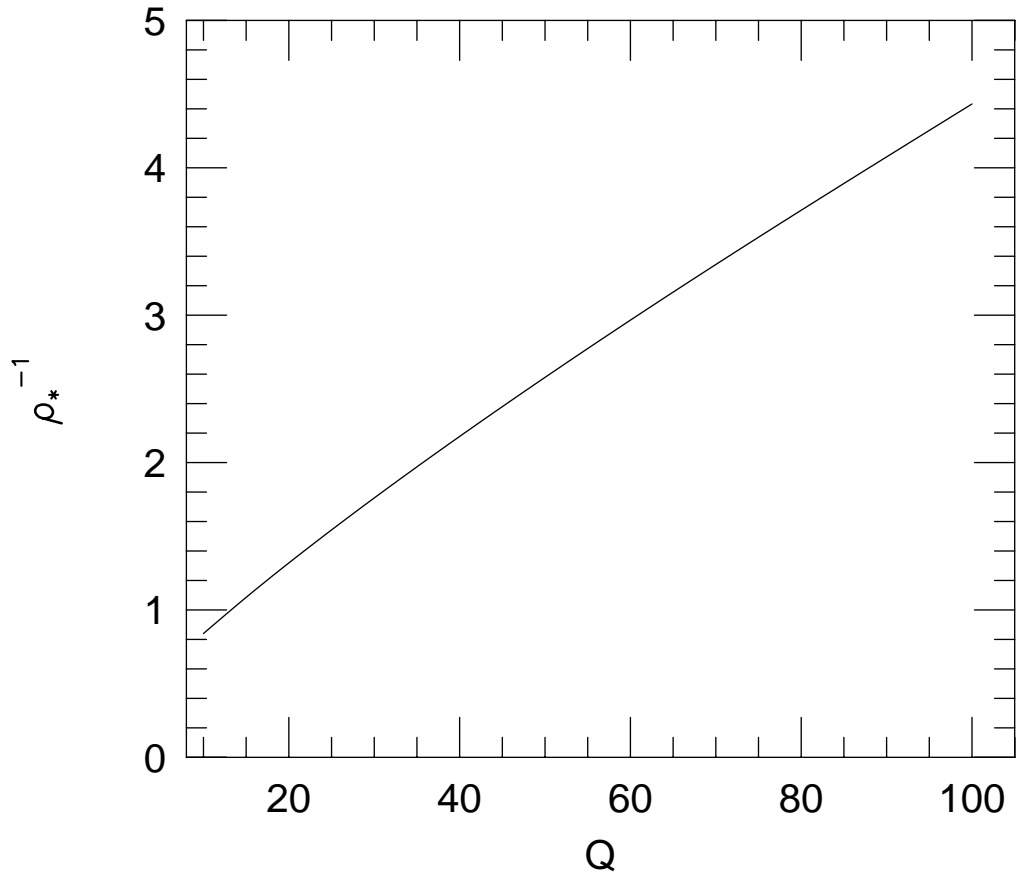
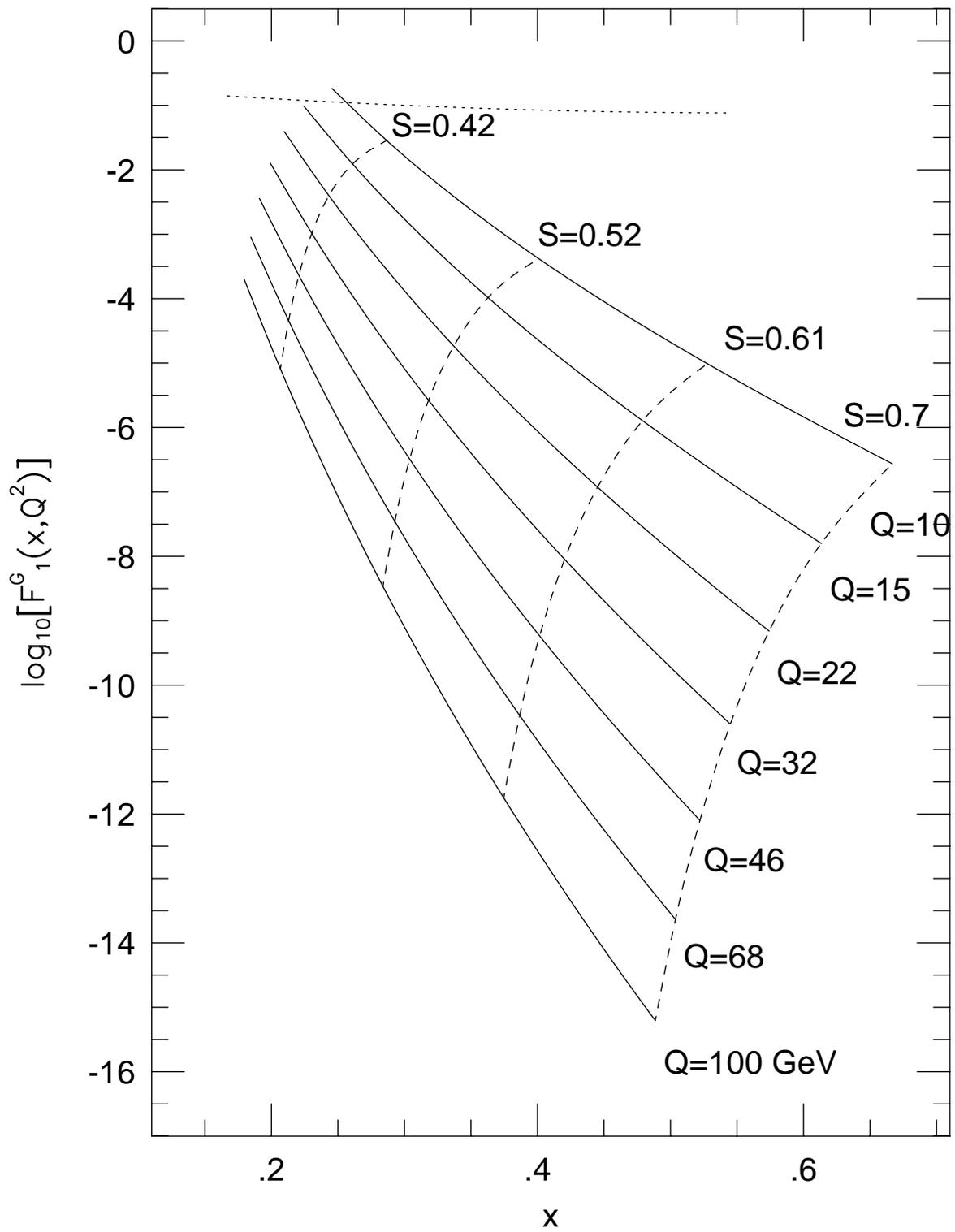
\end{document}